 \documentclass[manuscript]{emulateapj}
 \slugcomment{to appear in the Astrophysical Journal}
\shortauthors{Kato et al.}

 \usepackage{apjfonts}

\def\swift{{\it Swift~}}

\def\nova{{M31N~2008-12a~}}
\def\novak{{M31N~2008-12a}}

\newcommand{\cts}[1]{$\times 10^{#1}$ ct s$^{-1}$}

\newcommand{\power}[1]{$10^{#1}$}

\begin{document}

\title{X-RAY FLASHES IN RECURRENT NOVAE:\\
M31N 2008-12\lowercase{a}
AND THE IMPLICATIONS OF THE {\textit{Swift}} NON-DETECTION}

\author{Mariko Kato} 
\affil{Department of Astronomy, Keio University, Hiyoshi, Yokohama
  223-8521, Japan;}
\email{mariko.kato@hc.st.keio.ac.jp}

\author{Hideyuki Saio}
\affil{Astronomical Institute, Graduate School of Science,
    Tohoku University, Sendai, 980-8578, Japan}

\author{Martin Henze}
\affil{Institut de Ci\`encies de l'Espai (CSIC-IEEC), Campus UAB, C/Can Magrans s/n, 08193 Cerdanyola del Valles, Spain}

\author{Jan-Uwe Ness}
\affil{European Space Astronomy Centre, P.O.\ Box 78, 28691 Villanueva de la Ca\~{n}ada, Madrid, Spain}

\author{Julian P. Osborne, Kim L. Page}
\affil{X-Ray and Observational Astronomy Group, Department of Physics \& Astronomy, University of Leicester, LE1 7RH, UK}

\author{Matthew J. Darnley, Michael F. Bode}
\affil{Astrophysics Research Institute, Liverpool John Moores University, IC2 Liverpool Science Park, Liverpool, L3 5RF, UK}

\author{Allen W. Shafter}
\affil{Department of Astronomy, San Diego State University, San Diego, CA 92182, USA}

\author{Margarita Hernanz}
\affil{Institut de Ci\`encies de l'Espai (CSIC-IEEC), Campus UAB, C/Can Magrans s/n, 08193 Cerdanyola del Valles, Spain}

\author{Neil Gehrels}
\affil{Astrophysics Science Division, NASA Goddard Space Flight Center, Greenbelt, MD 20771, USA}

\author{Jamie Kennea}
\affil{Department of Astronomy and Astrophysics, 525 Davey Lab, Pennsylvania State University, University Park, PA 16802, USA}

\and
\author{Izumi Hachisu}
\affil{Department of Earth Science and Astronomy, College of Arts and
Sciences, The University of Tokyo, 3-8-1 Komaba, Meguro-ku,
Tokyo 153-8902, Japan}

\begin{abstract} 
Models of nova outbursts suggest that 
an X-ray flash should occur just after hydrogen ignition.  
However, this X-ray flash has never been observationally confirmed.
We present four theoretical light curves of the X-ray flash for 
two very massive  white dwarfs (WDs) of 1.380 and 1.385 $M_\sun$ and for two  
recurrence periods of 0.5 and 1 years. 
The duration of the X-ray flash is shorter for a more massive WD and 
for a longer recurrence period. 
The shortest duration of 14 hours (0.6 days) among the four cases is obtained
for the $1.385~M_\sun$ WD with one year recurrence period. 
In general, a nova explosion is relatively weak for a very short recurrence
period, which results in a rather slow evolution toward the optical peak.  
This slow timescale and the predictability of very short 
recurrence period novae give us a chance to observe X-ray flashes
of recurrent novae.
In this context, we report the first attempt, 
using the \swift observatory, to detect an X-ray flash
of the recurrent nova M31N 2008-12a (0.5 or 1 year recurrence period), 
which resulted in the non-detection of X-ray emission during 
the period of 8 days before the optical detection. 
We discuss the impact of these observations on nova outburst theory.
The X-ray flash is one of the last frontiers of nova studies and 
its detection is essentially important to understand the pre-optical-maximum
phase.  We encourage further observations.
\end{abstract}


\section{INTRODUCTION} \label{sec_introduction}

\begin{figure}
\epsscale{1.10}
\plotone{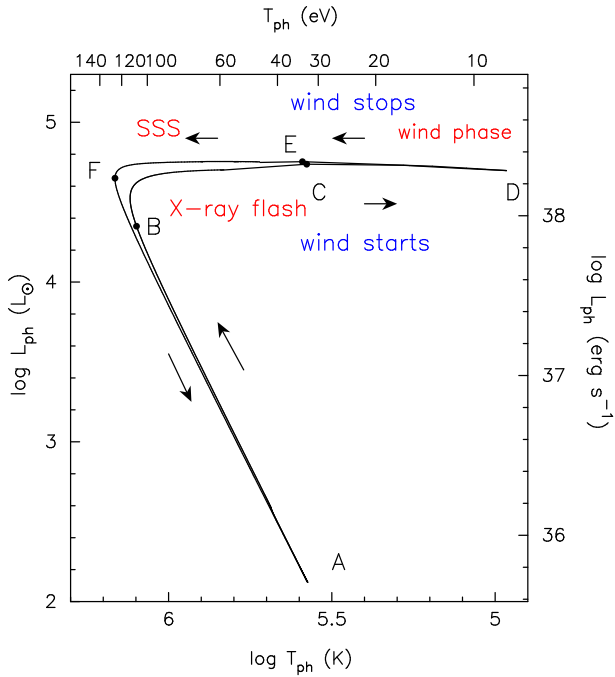}
\caption{Schematic HR diagram for one cycle 
of a nova outburst on a $1.38~M_\sun$ WD. 
A mass-accreting WD stays at point A. 
When unstable hydrogen burning sets in, the star becomes bright (goes up).
Point B denotes the epoch of maximum nuclear luminosity. 
Then the envelope expands and the photospheric temperature decreases with time
(goes rightward). 
The optically thick wind starts at point C. The photospheric radius reaches 
maximum at point D. 
A part of the envelope matter is blown away in 
the wind. The optically thick wind continues until point E. 
Hydrogen nuclear burning extinguishes at point F. 
Finally the star cools down to point A. 
Three stages, X-ray flash (from B to C), wind phase (from C to E through D), 
and supersoft X-ray phase (from E to F) are indicated.
}
\label{hr}
\end{figure}

A nova is a thermonuclear runaway event that occurs on an accreting 
white dwarf (WD) \citep[e.g.,][]{ibe82,jos93,nar80,pri95,sta74}. 
Figure \ref{hr} shows a schematic HR diagram for one cycle of a nova outburst 
on a very massive WD. The thermonuclear runaway of hydrogen sets in on 
an accreting WD at point A. The luminosity increases toward point B
at which the nuclear luminosity ($L_{\rm nuc}$) reaches its maximum.
After that, the envelope on the WD greatly expands and reaches 
point D (the maximum expansion of the photosphere:
corresponding to the optical peak). 
An optically thick wind begins to blow at point C and continues 
until point E through D. 
A part of the envelope mass is lost in the wind. 
From point C to E, the hydrogen-rich envelope mass decreases
owing to wind mass loss and nuclear burning. 
After point E, it decreases owing to hydrogen burning. 
The hydrogen burning extinguishes at point F.

The decay phase of optical and near-infrared (NIR) light curves corresponds 
to the phase from point D to E. 
The supersoft X-ray phase corresponds to the phase from point E to F. 
These phases have been well observed in a number of novae
in various wavelength bands 
\citep[e.g.][and references therein]{hac06kb,hac10,hac14k,hac15k,hac16k,osb15,sch11}. 
The evolution of novae has been modeled by 
the optically thick wind theory \citep{kat94h}, and their theoretical 
light curves for D-E-F  have successfully reproduced the observed light curves 
including NIR, optical, UV, and supersoft X-rays.  
From point D to E, the optical/IR light curves 
are well explained in terms of free-free emission \citep{gal76}, 
the fluxes of which are 
calculated from the mass-loss rate of the optically thick winds \citep{wri75}. 
From point E to F, the duration of the
supersoft X-ray phase is theoretically reproduced.
Detailed comparison with theory and observation enables us to 
determine/constrain the nova parameters such as the WD mass,
distance, and extinction, in many novae \citep{hac14k,hac15k,hac16k,hac16II}. 
Thus, the characteristic properties of a nova from D to F 
have been well understood in both observational and theoretical terms. 

The X-ray flash is the stage from point B to C, 
which occurs just after the hydrogen ignition \citep{kat15sh, hac16sk}, 
but {\it before} the optical discovery.
This stage has not been theoretically studied well, partly because of 
numerical difficulties and partly because of insufficient observational
data to guide the theoretical models.
In general, we cannot know in advance when and where a nova will erupt. 
Thus, soft X-ray flashes have never been detected in any kind of nova 
with any X-ray satellite. 
X-ray flashes represent one of the last frontiers of nova eruption studies 
and their detection will open a new landscape of nova physics. 

The X-ray flash of novae has been predicted from theoretical models for
many years 
\citep[e.g.,][]{sta90,kra02}, but its observation had not been 
attempted until recently.  
In an attempt to provide observational constraints on X-ray flashes
\citet{mor16} analyzed MAXI/GSC (Gas Slit Camera) data obtained with 
92 minute cadence for 40 novae, including recurrent novae. 
They deduced the upper limit of the soft X-ray fluxes 
spanning a period of 10 days before the optical discovery of each nova.  
The energy bandpass of MAXI/GSC, however, is too high (2-4 keV) to detect
the supersoft X-rays \citep[blackbody temperatures up to a maximum
of 120~eV, observed in nova \novak, see][]{2014A&A...563A...2H,hen15} expected during the flash.
Thus their upper limits of the bolometric luminosity 
are much higher than the theoretically expected values 
\citep[$\sim 10^{38}$ erg~s$^{-1}$, see][]{kat15sh} and
their approach was not effective to restrict the epoch of an X-ray flash. 

We carried out a coordinated, very high-cadence observing campaign with the 
\swift satellite \citep{2004ApJ...611.1005G} to detect the X-ray flash 
during the 2015 outburst of the recurrent nova \nova 
\citep{dar14,dar15,hen14,hen15,tan14}.  
This is the \textit{ideal} object to detect X-ray flashes
because its recurrence period is as short as a year, 
possibly even half a year \citep{hen15.half.period}. 
Such a very short recurrence period allows us to predict the eruption date 
with unprecedented accuracy ($\pm1$~month) and thereby makes any observational 
campaigns significantly more feasible than for any other novae. 
We found no significant X-ray emission during the eight days before 
the optical discovery by \citet{2015ATel.7964....1D}. This 
result is not consistent with the prediction made by \citet{kat15sh},
and suggests that theoretical models are still incomplete
especially in the rising phase.  Because no observational detection 
of soft X-rays and their properties has ever been obtained 
in the pre-optical-maximum phase,
we are unable to constrain the theoretical models.
In the present paper we describe the theoretical light curves of X-ray flashes
for massive WDs, and present the observational results. We also address the  
implication of a non-detection of a flash. 

This paper is organized as follows.  Section \ref{section_model} describes
our improved numerical calculations and presents theoretical
light curves of X-ray flashes as well as the physical properties of
expanding envelopes in the early phase of shell flashes. 
Section \ref{section_observation} describes the 
{\it Swift} observations of the 2015 outburst of M31N 2008-12a, which
resulted in the non-detection of an X-ray flash.
In Section \ref{section_implication},
we identify the reason why X-ray flash emission was not detected. 
Discussion and conclusion follow in Sections \ref{section_discussion}
and \ref{section_conclusion}.


\section{EARLY EVOLUTION OF SHELL FLASH}\label{section_model}

\subsection{Numerical method}\label{section_method}

We calculated recurrent nova models on 1.38 and 1.385 $M_\sun$ WDs
accreting hydrogen-rich matter ($X=0.70,~Y=0.28,$ and $Z=0.02$ for hydrogen, 
helium, and heavy elements, respectively) with mass-accretion rates of 
1.4 -- 2.5 $\times 10^{-7}~M_\sun$~yr$^{-1}$, 
corresponding to recurrence periods from one year to half a year 
\citep{dar15,hen15.half.period}.
We also calculated models for a 1.35 $M_\sun$ WD of 1.0 and 12 year 
recurrence periods for comparison. 
Table \ref{table_models} summarizes our models.  
The WD mass and mass-accretion rate are the given model parameters 
and the other values, including recurrence periods, 
are calculated results.

We calculated several outburst cycles until the shell flashes 
reached a limit cycle. 
We used the same Henyey-type code as in \citet{kat14shn,kat15sh}
and \citet{hac16sk}, but adopted a thinner static boundary layer 
[$2\times (M_{\rm WD}/M_\sun) \times 10^{-10}~M_\sun$]
for the outermost surface layer and smaller time-steps and mass zones. 
These technical improvements enabled us to calculate the photospheric values 
much more accurately in the extended phase (after point B in Figure \ref{hr}) 
until the optically thick wind begins
to blow at $\log T_{\rm ph}$ (K) $\sim 5.5$. The X-ray light curves
are calculated from the photospheric luminosity $L_{\rm ph}$
and temperature $T_{\rm ph}$ assuming blackbody emission.  

Our report here is focused on the very early phase of
nova outbursts, i.e., the X-ray flash phase. 
The occurrence of the optically thick wind in our models is judged 
using the surface boundary condition BC1 
listed in Table A1 of \citet{kat94h}.


\begin{figure}
\epsscale{1.10}
\plotone{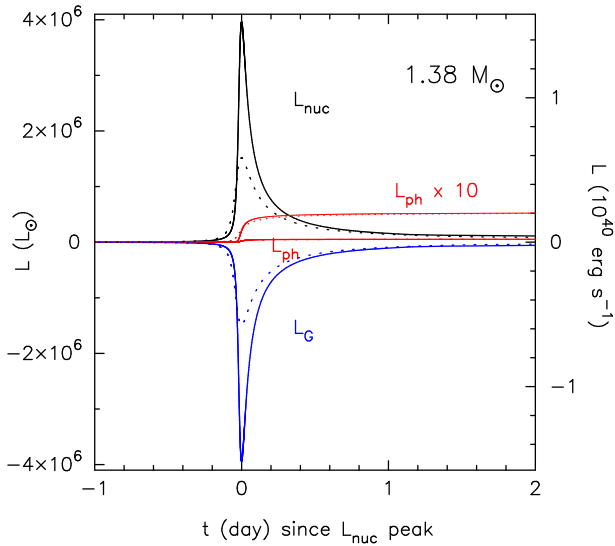}
\caption{
Evolution of the nuclear burning luminosity, $L_{\rm nuc}$, 
photospheric luminosity,  $L_{\rm ph}$, and gravitational energy release
rate, $L_{\rm G}$, of a shell flash on a $1.38~M_\sun$ WD.
A large amount of nuclear luminosity is produced but is absorbed
in the burning shell as expressed by a large negative value of $L_{\rm G}$.
As a result, the outward radiative luminosity, i.e., the photospheric 
luminosity, $L_{\rm ph}$, is very small. 
Thin lines denote 10 times the photospheric luminosity, 
$L_{\rm ph} \times 10$.
Solid lines denote those for a $P_{\rm rec}=0.95$ yr model while
dotted lines represent those for a $P_{\rm rec}=0.47$ yr model.
}
\label{L.138}
\end{figure}

\subsection{Energy budget}
Figure \ref{L.138} shows the energy budget in the very early phase of 
shell flashes on a $1.38~M_\sun$ WD for two recurrence periods 
$P_{\rm rec}=0.95$~yr ($\dot M_{\rm acc}=1.6 \times 10^{-7}~M_\sun$yr$^{-1}$: 
solid) and  $P_{\rm rec}=0.47$~yr ($\dot M_{\rm acc}=2.5 
\times 10^{-7}~M_\sun$yr$^{-1}$: dashed). 
The nuclear luminosity, 
\begin{equation}
L_{\rm nuc} = \int_0^M \epsilon_n d M_r, 
\end{equation}
takes a maximum value of $L_{\rm nuc}^{\rm max}
=3.9 \times 10^6~L_\sun$ for the $P_{\rm rec}=0.95$ yr
case and $L_{\rm nuc}^{\rm max}= 1.5 \times 10^6~L_\sun$
for the $P_{\rm rec}=0.47$ yr case.
Here, $\epsilon_n$ is the energy generation rate per unit mass
for hydrogen burning, $M_r$ is the mass within the radius $r$, and $M$ is the 
mass of the white dwarf including the envelope mass.
The maximum value is lower for the shorter recurrence period. 
A shorter recurrence period corresponds to a higher mass-accretion rate, 
and ignition starts at a smaller envelope mass because of heating by a larger 
gravitational energy release rate.  
We define the hydrogen-rich envelope mass 
as the mass above $X=0.1$, i.e.,
\begin{equation}
M_{\rm env} = \int_{X>0.1} d M_r, 
\end{equation}
and the ignition mass, $M_{\rm ig}$, as the hydrogen-rich 
envelope mass at the maximum nuclear energy release rate, i.e., 
$M_{\rm ig}= M_{\rm env}$ (at $L_{\rm nuc}=L_{\rm nuc}^{\rm max}$)
because the envelope mass is
increasing owing to mass-accretion even after hydrogen ignites.
The ignition mass is $2.0 \times 10^{-7}~M_\sun$ for $P_{\rm rec}=0.95$ yr
and $1.6 \times 10^{-7}~M_\sun$ for $P_{\rm rec}=0.47$ yr.
For a smaller envelope mass, the pressure at the bottom of the envelope
is lower and therefore the maximum temperature is also lower. 
As a result, the maximum value of $L_{\rm nuc}^{\rm max}$ is lower for a 
shorter recurrence period.

Although a high nuclear luminosity ($\sim 10^6~L_\sun$) is produced, 
most of the energy is absorbed by the burning shell, as indicated by the
large negative values of the gravitational energy release rate, 
$L_{\rm G}$, 
which is defined by
\begin{equation}
L_{\rm G} = \int_0^M \epsilon_g d M_r = \int_0^M -T \left({{\partial 
s} \over {\partial t}}\right)_{M_r} d M_r, 
\end{equation}
where $\epsilon_g$ is the gravitational energy release rate per unit mass,
$T$ is the temperature and $s$ is the entropy per unit mass 
\citep[see, e.g.,][]{kat14shn, hac16sk}.

As a result, the photospheric luminosity $L_{\rm ph}$ 
($\approx L_{\rm nuc}-|L_{\rm G}|$, because neutrino loss is negligible) 
is two orders of magnitude smaller than the peak value of 
$L_{\rm nuc}^{\rm max}$. 
Shortly after the ignition, 
the photospheric luminosity approaches a constant value. 
This constant value is close to but slightly smaller than 
the Eddington luminosity at the photosphere, 
\begin{equation}
L_{\rm Edd,ph} = {{4\pi cG{M_{\rm WD}}} \over {\kappa_{\rm ph}}}
= 2.0 \times 10^{38} {\rm ~erg~s}^{-1}
\left({{M_{\rm WD}}\over {1.38~M_\sun}}\right)
\left({{0.35} \over {\kappa_{\rm ph}}}\right),
\label{equation_Edd_ph}
\end{equation}
where $\kappa_{\rm ph}$ is the opacity at the photosphere.
In other words, the photospheric luminosity stays below the Eddington luminosity 
in this early phase of a shell flash.


\begin{figure}
\epsscale{1.10}
\plotone{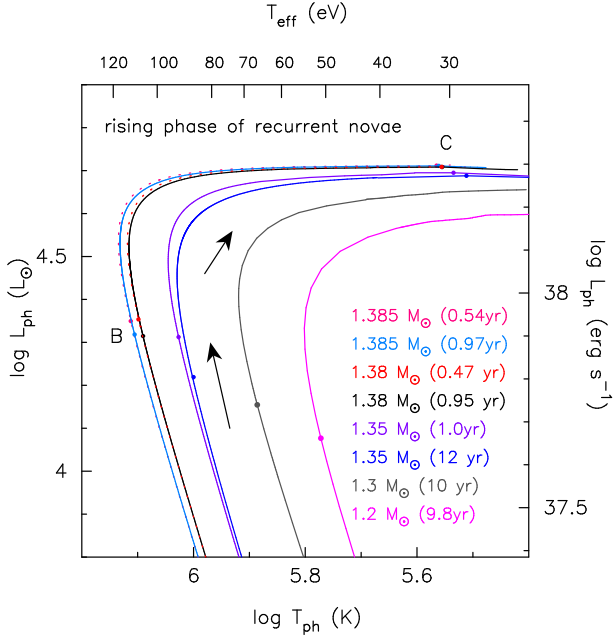}
\caption{
HR diagram for the rising phase of recurrent novae.  
The WD mass and recurrence period are indicated by different colors. 
The maximum nuclear luminosity at point B, 
$L_{\rm nuc}=L_{\rm nuc}^{\rm max}$, and the
occurrence of the optically thick wind mass-loss at point C are indicated by 
the small filled circles.  For less massive WDs ($\le 1.3~M_\sun$)
point C is located at $\log T_{\rm ph}$ (K) $<5.4$, beyond the 
right edge of the figure. 
The dotted lines indicate the shorter recurrence period models of 1.38 and
1.385 $M_\sun$.} 
\label{hr.rising}
\end{figure}

\subsection{HR diagram}

Figure \ref{hr.rising} shows the HR diagram of the rising phase of 
recurrent nova outbursts for various WD masses and recurrence periods. 
X-ray flashes correspond to the phase approximately from point B to C 
(the same marks denote the same stage in Figure \ref{hr}). 
A more massive WD reaches a higher photospheric luminosity and maximum 
photospheric temperature, therefore we expect larger X-ray luminosity
during the X-ray flash on a more massive WD. 

The track in the HR diagram depends not only on the WD mass but also 
more weakly on the recurrence period. For a longer recurrence period, 
the ignition mass is larger and the envelope begins to expand at a lower 
luminosity. Thus, the track locates slightly lower 
and towards the right (redder) 
side compared to that of a shorter recurrence period. 


\begin{figure}
\epsscale{1.10}
\plotone{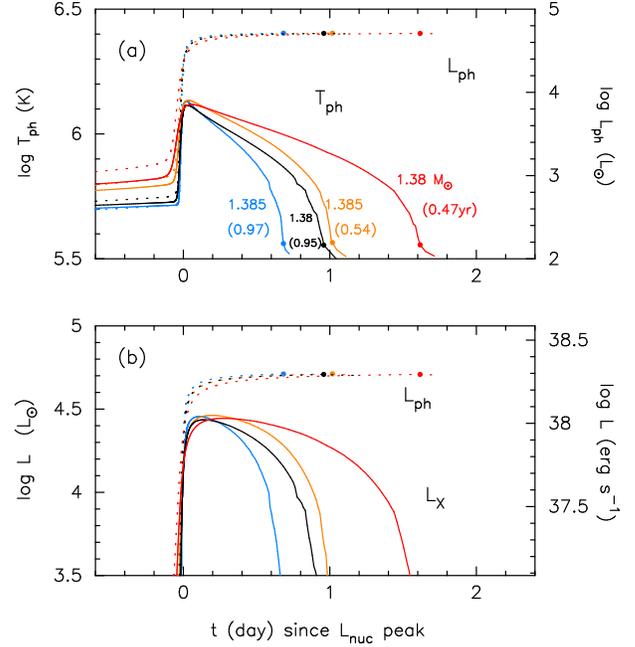}
\caption{
(a) Photospheric temperature $T_{\rm ph}$ (solid lines) 
and luminosity $L_{\rm ph}$ (dotted lines),
and (b) X-ray luminosity $L_{\rm X}$ (solid lines: 0.3--1.0 keV) 
and luminosity $L_{\rm ph}$ (dotted lines)
during X-ray flashes, against time after the ignition.
The origin of time $t=0$ is defined  
as the point B where $L_{\rm nuc}=L_{\rm nuc}^{\rm max}$.  
The blue lines denote the model of $1.385~M_\sun$ 
with $P_{\rm rec}=0.97$ yr, orange lines mark $1.385~M_\sun$
with $P_{\rm rec}=0.54$ yr, black lines mark $1.38~M_\sun$ 
with $P_{\rm rec}=0.95$ yr, and red lines mark $1.38~M_\sun$
with $P_{\rm rec}=0.47$ yr.  Point C is indicated by a dot,
but point C on the X-ray light curves are located
below the lower bound in panel (b). 
}
\label{flash.3}
\end{figure}

\subsection{X-ray light curve}
Figure \ref{flash.3}(a) shows the photospheric temperature and luminosity 
during the X-ray flashes for the 1.38 and 1.385 $M_\sun$ WD models 
in Table \ref{table_models}. 
The photospheric luminosity quickly rises near the $L_{\rm nuc}$ peak ($t=0$) 
and reaches a constant value. 
The photospheric temperature reaches its maximum immediately after 
the time of ignition and decreases with time. 
When the envelope expands and the photospheric temperature decreases to 
a critical temperature, the optically thick wind occurs (this 
epoch corresponds to point C in Figure \ref{hr}). 
This critical temperature is indicated by small filled circles in Figure \ref{flash.3}(a).
Shortly before this epoch, the temperature drops quickly corresponding 
to the opacity increase near the photosphere, which will be discussed 
in Section \ref{section_internal}.

Figure \ref{flash.3}(b) shows the X-ray luminosity 
in the supersoft X-ray band (0.3 -- 1.0 keV). 
The duration ($\log L_X/L_\sun >$ 4) of the X-ray flash is 14 -- 19 hours (0.59 -- 0.78 days) 
for $\sim1$ year recurrence period novae and 22 -- 34 hours (0.9 -- 1.4 days)
for $\sim0.5$-year period novae. 
For a shorter recurrence period, the ignition is weaker as explained before, 
so the expansion is slower than in longer recurrence period novae. 
The shortest duration among these four models is 14 hours (0.59 days)
for 1.385 $M_\sun$ with $P_{\rm rec}=0.97$ yr.
As 1.385 $M_\sun$ is almost the upper limit of a mass accreting WD
with no rotation \citep{nom84ty}, a duration of 14 hours (0.59 days) 
would be the minimum for novae with recurrence periods shorter 
than one year. 

The ultra-short recurrence period nova M31N 2008-12a shows
a supersoft X-ray source phase (SSS) of 10 days \citep{hen14,hen15,tan14}.  
In general, the SSS phase (from E to F in Figure \ref{hr}) 
is shorter for a more massive WD. 
The duration of the SSS phase of M31N 2008-12a is consistent with a 
$\sim 1.38~M_\sun$ WD \citep{hen15}. 
Such a SSS phase duration allows us to exclude WDs much more 
massive than 1.385 $M_\sun$. Similarly, we can also exclude a $1.35~M_\sun$ WD 
because its SSS duration would be too long.  The duration 
of the X-ray flash in a 1.38 -- $1.385~M_\sun$ WD ($> 14$ hours)
is long enough to be detectable with the 6-hour cadence of our {\it Swift} 
observations (see Section \ref{section_observation}).


\begin{figure}
\epsscale{1.10}
\plotone{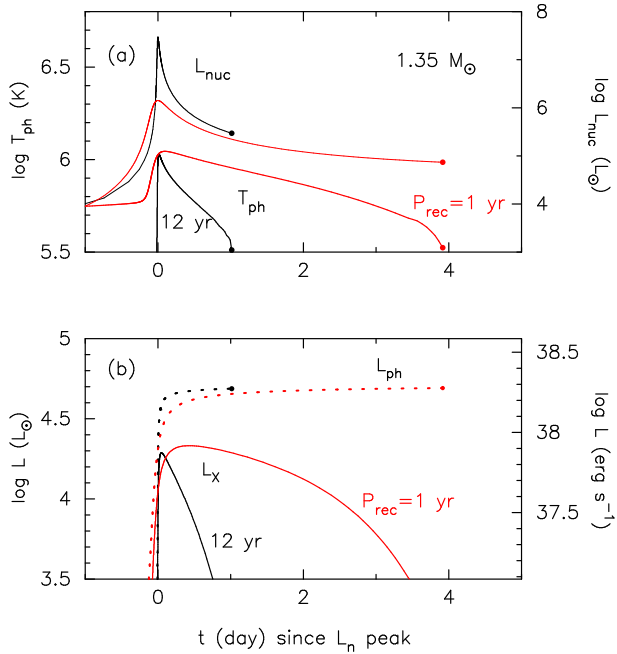}
\caption{
(a) Nuclear burning luminosity $L_{\rm nuc}$ (thin solid line) and photospheric 
temperature $T_{\rm ph}$ (thick solid line) for our $1.35~M_\sun$ model with 
$P_{\rm rec}=12$ yr (black) and 1 yr (red).  
(b) The photospheric luminosity $L_{\rm ph}$ and X-ray luminosity
$L_{\rm X}$ for the same models as in panel (a). 
}
\label{flash.m135}
\end{figure}

\subsection{Various WD models and flash duration}
We have calculated shell flash models on a $1.35~M_\sun$ WD with 
$P_{\rm rec}=1$ and 12 yr for comparison. 
The corresponding mass accretion rates are listed in Table \ref{table_models}.
Figure \ref{flash.m135}(a) shows the evolution of the 
nuclear luminosity and photospheric temperature. 
The outburst in the $P_{\rm rec}=1$ yr case (red lines) is much weaker than  
in the $P_{\rm rec}=12$ yr scenario (black lines) as indicated
by the lower nuclear energy generation rate $L_{\rm nuc}$. 
As a result, in the $P_{\rm rec}=1$ yr case, the photosphere slowly expands,
therefore the photospheric temperature decreases slowly, 
which results in a much longer X-ray flash 
as demonstrated in Figure \ref{flash.m135}(b).  


\begin{deluxetable}{llcllcc}
\tabletypesize{\scriptsize}
\tablecaption{Summary of Recurrent Nova Models
\label{table_models}}
\tablewidth{0pt}
\tablehead{
\colhead{WD mass} &
\colhead{  } &
\colhead{$\dot M_{\rm acc}$} &
\colhead{$t_{\rm rec}$} &
\colhead{$t_{\rm X-flash}^a$}&
\colhead{$L_{\rm nuc}^{\rm max}$} &
\colhead{$ T_{\rm nuc}^{\rm max}$}\\
\colhead{ ($M_\odot$) } &
\colhead{  } &
\colhead{($10^{-7}M_\odot$ yr$^{-1}$)   } &
\colhead{ (year)}& 
\colhead{(day)}&
\colhead{($ 10^6 L_\odot $) } &
\colhead{($10^8$ K)}
}
\startdata
1.385 & ... & $2.0 $ & 0.54  &0.90  & 2.3  & 1.74 \\
1.385 & ... & $1.4 $ & 0.97  &0.59 & 5.1  & 1.84 \\
1.38 & ... & $2.5 $ & 0.47  &1.4  & 1.5   & 1.66 \\
1.38 & ... & $1.6 $ & 0.95  &0.78 & 3.9  & 1.77 \\
1.35 & ... & $2.6 $ & 1.0  &2.5  & 1.5   & 1.54 \\
1.35 & ... & $0.5 $ & 12  & 0.37   & 29   & 1.89
\enddata
\tablenotetext{a}{Duration of the X-ray flash: $L_{\rm X}$(0.3 - 1.0 keV) $> 10^{4}~L_\sun$.}
%
\end{deluxetable}

Table \ref{table_models} lists the maximum value of the nuclear energy
generation rate $L_{\rm nuc}^{\rm max}$ and the maximum temperature
in the hydrogen nuclear burning region $T_{\rm nuc}^{\rm max}$.
There are three models of similar recurrence period, 
$P_{\rm rec} \sim 1$ year, for 1.35, 1.38 and 1.385 $~M_\sun$.   
Both $T_{\rm nuc}^{\rm max}$ and $L_{\rm nuc}^{\rm max}$ are
larger in more massive WDs. 
This means that the shell flash is stronger and hence evolves faster in 
a more massive WD with the same recurrence period
because of the stronger gravity of the WD. 
On the other hand, for a given WD mass, both $T_{\rm nuc}^{\rm max}$ and 
$L_{\rm nuc}^{\rm max}$ are smaller 
for a shorter recurrence period. 
This tendency is clearly shown in the two $1.35~M_\sun$ models  
in which $L_{\rm nuc}^{\rm max}$ is 19 times larger in $P_{\rm rec}=12$ yr
than in $P_{\rm rec}=1$ yr.
The duration of the X-ray flash is 6.8 times longer
for the shorter recurrence period.

To summarize, more massive WDs undergo stronger shell flashes,
but their flashes become weaker for shorter recurrence periods. 
Thus, the duration of X-ray flash is shorter in more 
massive WDs, but longer for shorter recurrence periods. 
Even in WDs as massive as 1.38 $~M_\sun$, the X-ray flash 
could last $\sim 0.5$ days.


\begin{figure}
\epsscale{1.10}
\plotone{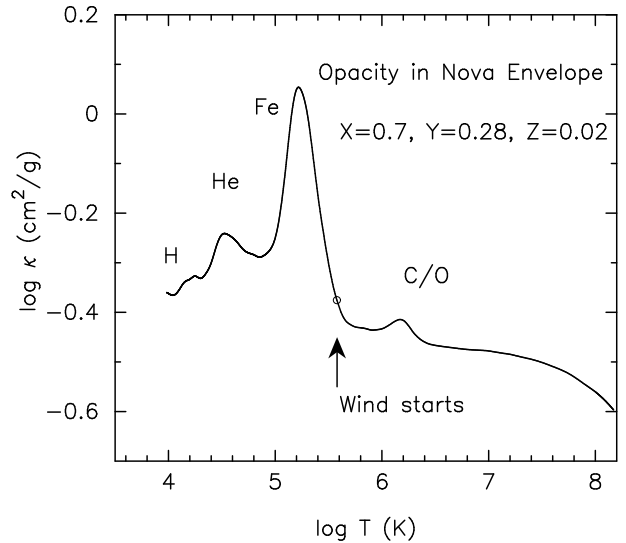}
\caption{
Example for the OPAL opacity run for $X=0.7, Y=0.28$, and $Z=$0.02.
The opacity is taken from \citet{igl96} for the structure of 
an extended wind solution with $\log T_{\rm ph}$ (K) = 4.0
on the $1.38 M_\odot$ WD.
The small open circle labeled ``Wind starts'' denotes the critical point.
The main element responsible for each opacity peak is indicated 
by its atomic symbol. When a nova envelope expands and the photospheric 
temperature decreases to $\log T$ (K) $ \sim 5.5$, optically thick winds 
are accelerated owing to the large Fe peak.   
See the main text for more details.
}
\label{opac}
\end{figure}

\subsection{Internal structure at the end of X-ray flash}
\label{section_internal}

The X-ray flash ends when the envelope expands and the 
optically thick winds start blowing. 
In this subsection, we examine the possibility 
that the optically thick winds are accelerated much earlier
(i.e., before point C in Figure \ref{hr}), 
which shortens the duration of the X-ray flash.

Before going into the details of the envelope structure, 
it would be instructive to discuss the opacity in the envelope, 
which is closely 
related to the envelope expansion and occurrence of wind mass loss. 
Figure \ref{opac} shows the run of the OPAL opacity \citep{igl87,igl96} 
with solar composition 
in an optically thick wind solution with $\log T_{\rm ph}$ (K) =4.0 
on the $1.38~M_\odot$ WD. 
This model has a very large envelope mass and a uniform chemical 
composition that does not exactly correspond to the structure 
of a very short recurrence period nova, but is sufficient to show 
the characteristic properties of the OPAL opacity. 
In our evolution calculation, the chemical composition varies 
from place to place and the photospheric temperature is much higher 
than in this case. 

The opacity has several peaks above the constant value of
the electron scattering opacity 
$\log \kappa_{\rm el} =\log [0.2(1+X)]= \log 0.34 = -0.47$ for $X=0.7$.
The peak at $\log T$ (K) =4.5 corresponds to the second helium ionization.
The prominent peak at $\log T$ (K) $\sim 5.2$ is owed mainly to
low/mid-degree ionized iron found in opacity projects
\citep{igl87,sea94}. Hereafter, we call it the ``Fe peak.''
The peak at $\log T$ (K)= 6.2 relates to highly ionized Fe, C, O, and Ne.
We call it the ``C/O peak.'' 
A tiny peak around $\log T$ (K)= 7.0 is owed to the highly ionized
heavy elements Ar -- Fe. 
The opacity is smaller than that of electron
scattering at the highest temperature region because of the Compton effect.

A large peak in the opacity causes the envelope expansion and 
accelerates the optically thick winds.  In the model in Figure \ref{opac}, 
the critical point of the optically thick winds 
\citep{kat94h}, 
in which the velocity becomes equal to the isothermal, sound velocity,
appears just on the inside of the Fe peak. 
A critical point appears in the region of acceleration which means 
that the envelope is accelerated outward
where the opacity quickly increases outward.


\begin{figure}
\epsscale{1.10}
\plotone{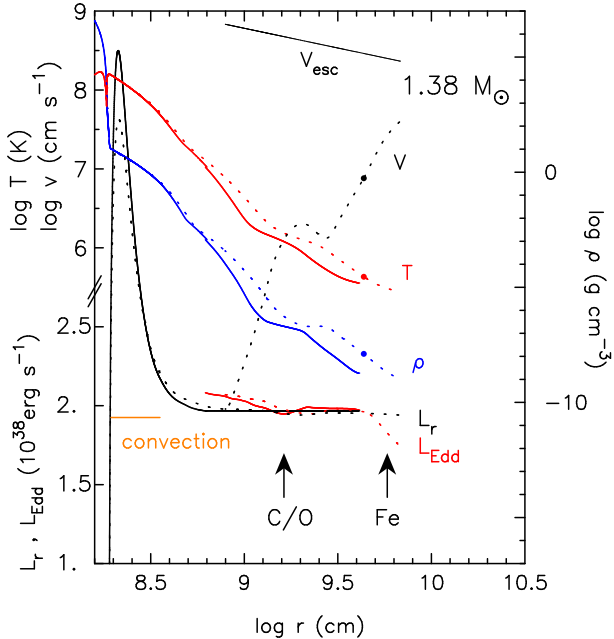}
\caption{
Envelope structures in two stages near point C for the evolution models of
$M_{\rm WD}=1.38~M_\odot$ with $P_{\rm rec}=0.95$ yr. 
From upper to lower, the escape velocity $V_{\rm esc}=\sqrt{2GM_{\rm WD}/r}$, 
wind velocity $V$, temperature $T$, density $\rho$, radiative luminosity 
$L_r$ which is the summation of diffusive luminosity 
and convective luminosity, 
and the local Eddington luminosity
$L_{\rm Edd}=4 \pi c G M_{\rm WD}/ \kappa$.
The position of the critical point \citep{kat94h}
is indicated by a small filled circle.
Two arrows indicate the regions corresponding to the C/O and Fe opacity peaks, 
respectively. The convective region is indicated by the horizontal orange line.
Solid lines denote the model at point C in Figure \ref{hr.rising} 
($\log T_{\rm ph}$ (K) =5.56). 
This is the solution just before the wind starts,
so no velocity profile appears. 
Dotted lines represent the model shortly after point C 
at $\log T_{\rm ph}$ (K)= 5.44.
}
\label{struc.startML}
\end{figure}

Figure \ref{struc.startML} shows the internal structures
at the end of the X-ray flash in our $1.38~M_\sun$ WD model with 
$P_{\rm rec}=0.95$ yr. 
The solid line represents the structure at point C in Figures \ref{hr},
\ref{hr.rising}, and \ref{flash.3}, and the dotted line corresponds
to the stages shortly after point C. 
From top to bottom we show the escape velocity $\sqrt{2GM_{\rm WD}/r}$,
wind velocity $V$, temperature $T$, density $\rho$, 
radiative luminosity $L_{\rm r}$ which is the summation of 
diffusive luminosity and convective luminosity, 
and local Eddington luminosity defined by
\begin{equation}
L_{\rm Edd} = {4\pi cG{M_{\rm WD}} \over\kappa},
\label{equation_Edd}
\end{equation}
where $\kappa$ is the opacity. 
As the opacity depends on the local temperature and density, the local 
Eddington luminosity also varies. Note that Equation (\ref{equation_Edd_ph})
represents the photospheric value of Equation (\ref{equation_Edd}). 
The velocity is not plotted for the solution at point C. 

The local Eddington luminosity in Figure \ref{struc.startML} 
shows a small dip at $\log r$ (cm) $\sim 9.2$ corresponding to the C/O peak
at $\log T~({\rm K})=6.2$ in Figure \ref{opac}. 
Here, the local Eddington luminosity is slightly lower than the 
diffusive luminosity, i.e., locally the luminosity is 
super-Eddington.  However, this C/O peak does not result 
in the occurrence of optically thick winds.  Instead, the temperature 
and density profiles become shallower in this region.  

When the envelope expands enough and the photospheric temperature 
approaches the prominent Fe peak, optically thick winds occur. 
The Fe peak is so large that the local Eddington luminosity decreases to much 
below the radiative luminosity. 
The critical point \citep{kat94h} appears  near the photosphere, 
which corresponds to the inner edge of the Fe peak in Figure \ref{opac}. 

If the winds were accelerated by the C/O peak, 
the X-ray flash durations would be 
much shorter because the expansion and acceleration occur much earlier. 
We have confirmed in all of our calculated models that the wind 
is driven by the Fe peak and not by the C/O peak. 
Thus, we conclude that the X-ray flash should last at least a half  
day as in Table \ref{table_models} and could not be much shorter than that.

\section{SEARCH FOR THE X-RAY FLASH IN THE 2015 ERUPTION OF M31N~2008-12a}
\label{section_observation}

\subsection{Observing Strategy}
The multiwavelength coverage of the 2013 and 2014 eruptions of \nova 
\citep{dar14,dar15,hen14,hen15} resulted in significantly improved 
predictions of future eruptions. Moreover, \citet{hen15.half.period} 
combined new findings with archival data to arrive at a $1\sigma$ prediction 
accuracy of $\pm 1$~month (and suggest a recurrence period of $175\pm11$ days). 
Based on the updated forecast we designed an observational campaign 
to monitor the emerging 2015 eruption and catch the elusive X-ray flash.

The project was crucially reliant on the unparalleled scheduling 
flexibility of the \swift satellite \citep{2004ApJ...611.1005G}, whose 
X-ray telescope \citep[XRT;][]{2005SSRv..120..165B} provided 
a high-cadence monitoring. Similarly, the unprecedented short 
recurrence time and predictability of \nova made it the only target 
for which such an endeavor was feasible.

Starting from 2015 August 20 UT, a 0.6~ks \swift XRT observation was 
obtained every six hours. After the first week of the monitoring campaign,
the exposure time per observation was increased from 0.6~ks to 1~ks, 
because the actual exposure time often fell short of the goal. 
The nova eruption was discovered on August 28 \citep{2015ATel.7964....1D},
slightly earlier than predicted by \citet{hen15.half.period},
without any prior detection of an X-ray flash. Because of this early eruption
date and the last-minute improvement in prediction accuracy, based on the
recovery of the 2010 eruption \citep{hen15.half.period}, only eight days worth
of observations were obtained before the 2015 eruption.
All individual observations until after the optical discovery are listed
in Table\,\ref{tab:obs}. The campaign continued until the end of the SSS
phase and the analysis of the phase is presented by Darnley et al. 
(2016, in prep.).

\subsection{Data Analysis}
All \swift XRT data were obtained in photon counting (PC) mode and were reduced
using the standard \swift and Heasarc tools (HEASOFT version 6.16). 
Our analysis started from the cleaned level 2 files that had been 
reprocessed locally with HEASOFT version 6.15.1 at the \swift UK data centre.

We extracted source and background counts in \texttt{xselect} v2.4c based 
on the XRT point spread function (PSF) of \nova observed during previous 
eruptions. We applied the standard grade selection 0--12 for PC mode
observations. Based on the early SSS phase detections we chose a circular 
region with a radius of 22 arcsec, which corresponds to a 78\% PSF area (based 
on the merged detections of the 2013/4 eruptions), to optimize the ratio 
of source to background counts in the source region. The background region 
excluded the locations of nearby faint X-ray sources as derived 
from the merged data of the 2013 and 2014 eruption monitoring campaigns 
\citep[cf.][]{hen14,hen15}. All counts were restricted 
to the 0.3--1.0~keV band 
(refer to the X-ray spectrum of the eruption discussed in Darnley et al. 
2016, in prep.).

We checked for a source detection using classical Poisson statistics and 
determined $3\sigma$ count rate upper limits using the method of 
\citet{1991ApJ...374..344K}. The number of background counts were scaled 
to the source region size and corrected for the differences in exposure, 
derived from the XRT exposure map, 
between the regions for each individual observation. To improve the signal 
to noise ratio of the detection procedure the dataset was smoothed by a
two-observation wide boxcar function
to achieve a rolling $\sim12$~hour window. 
The added source and background counts were analyzed in the same way 
as the individual measurements. Note, that therefore successive upper
limits are not statistically independent.


\begin{figure*}
\epsscale{0.90}
\plotone{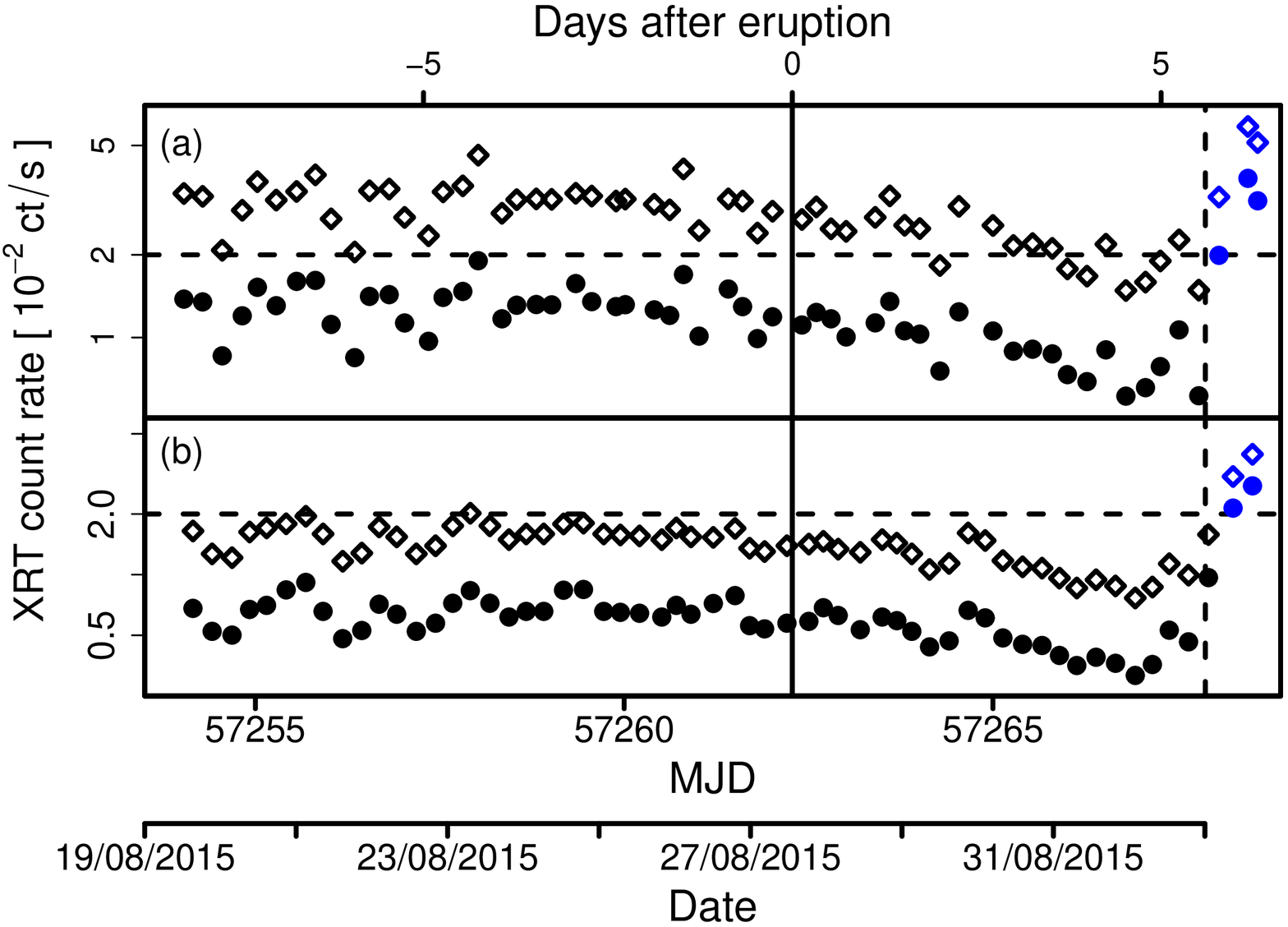}
\caption{
\swift XRT upper limits based on \citet{1991ApJ...374..344K} for the 
count rate of \nova assuming confidence levels $5\sigma$ (open diamonds) 
and $3\sigma$ (filled circles). Panels (a) and (b) show the individual 
and merged observations, respectively. The dashed horizontal line marks 
the expected XRT flash count rate \citep[cf.][]{hen15,kat15sh}. The solid vertical line indicates the 2015 
eruption date on August 28.28. The dashed vertical line marks the estimated onset
of the SSS phase (cf. Darnley et al. 2016, in prep.) and is followed by the blue data points
that indicate the formal upper limits corresponding to the early SSS detections, for comparison.
}
\label{fig:ulim}
\end{figure*}

\subsection{Results}
The monitoring campaign was executed exceptionally well, with a median 
cadence of 6.3~hours between two consecutive observations. At no point 
was there more than a 10~hour gap between successive pointings. 
Therefore, the minimum flash duration of 
14 hours (0.59~days) would have been covered 
by at least one observation, more likely two, during the entire eight days 
prior to the eruption.

In Figure \ref{fig:ulim} the resulting $3\sigma$ and $5\sigma$ XRT count 
rate upper limits are shown for the individual and merged observations, 
respectively. The $5\sigma$ upper limits are between 2 -- 4 \cts{-2} 
for the individual observations, which roughly corresponds to the 
variability range of the SSS phase around maximum \citep{hen15}. 
The combined $5\sigma$ upper limits for each two successive 
merged observations 
(i.e. a rolling $\sim$12~hour period) were almost entirely well below 
the expected flash count rate of 2\cts{-2}. This prediction assumes a
similar luminosity and spectrum for the X-ray flash and the SSS phase
\citep[see Fig.\,\ref{hr} and compare][]{hen15,kat15sh}.

The time of eruption ($T_E$) is defined as the midpoint between the last 
non-detection by the Liverpool Telescope (MJD 57262.16) and the first
\swift UVOT detection (MJD 57262.40; cf. Darnley et al. 2016, in prep.,
for both). Therefore, $T_E = \mbox{MJD}\,\, 57262.28 \pm0.12$ (August 28.28 UT), with the
error corresponding to half the interval between both observations.
The rightmost data points in 
Figure \ref{fig:ulim} feature the start of the SSS phase to compare 
the signatures of an actual detection. Additionally, the number of counts 
in the source region always remained below 2 for individual observations, 
except for the emergence of the SSS emission.


\begin{deluxetable*}{rrrrrrr}
\tabletypesize{\scriptsize}
\tablecaption{\swift observations for the 2015 X-ray flash monitoring of nova \novak.
\label{tab:obs}}
\tablewidth{0pt}
\tablehead{
\colhead{ObsID} & \colhead{Exp\tablenotemark{a}} & \colhead{Date\tablenotemark{b}} & \colhead{MJD\tablenotemark{b}} & \colhead{$\Delta t$\tablenotemark{c}} & \colhead{$3\sigma$ ulim} & \colhead{$5\sigma$ ulim}\\
\colhead{} & \colhead{[ks]} & \colhead{[UT]} & \colhead{[d]} & \colhead{[d]} & \colhead{[\power{-2} ct s$^{-1}$]} & \colhead{[\power{-2} ct s$^{-1}$]} }
\startdata
00032613063 & 0.66 & 2015-08-20.027 & 57254.0273 & -8.253 & $<1.4$ & $<3.3$ \\
00032613064 & 0.47 & 2015-08-20.281 & 57254.2812 & -7.999 & $<1.3$ & $<3.3$ \\
00032613065 & 0.70 & 2015-08-20.547 & 57254.5469 & -7.733 & $<0.9$ & $<2.1$ \\
00032613066 & 0.50 & 2015-08-20.820 & 57254.8203 & -7.460 & $<1.2$ & $<2.9$ \\
00032613067 & 0.40 & 2015-08-21.023 & 57255.0234 & -7.257 & $<1.5$ & $<3.7$ \\
00032613068 & 0.48 & 2015-08-21.281 & 57255.2812 & -6.999 & $<1.3$ & $<3.2$ \\
00032613069 & 0.53 & 2015-08-21.555 & 57255.5547 & -6.725 & $<1.6$ & $<3.4$ \\
00032613070 & 0.41 & 2015-08-21.812 & 57255.8125 & -6.467 & $<1.6$ & $<3.9$ \\
00032613071 & 0.55 & 2015-08-22.023 & 57256.0234 & -6.257 & $<1.1$ & $<2.7$ \\
00032613072 & 0.72 & 2015-08-22.344 & 57256.3438 & -5.936 & $<0.8$ & $<2.0$ \\
00032613073 & 0.54 & 2015-08-22.543 & 57256.5430 & -5.737 & $<1.4$ & $<3.4$ \\
00032613074 & 0.42 & 2015-08-22.812 & 57256.8125 & -5.467 & $<1.4$ & $<3.5$ \\
00032613075 & 0.54 & 2015-08-23.020 & 57257.0195 & -5.260 & $<1.1$ & $<2.7$ \\
00032613076 & 0.78 & 2015-08-23.344 & 57257.3438 & -4.936 & $<1.0$ & $<2.3$ \\
00032613077 & 0.45 & 2015-08-23.543 & 57257.5430 & -4.737 & $<1.4$ & $<3.4$ \\
00032613078 & 0.51 & 2015-08-23.809 & 57257.8086 & -4.471 & $<1.5$ & $<3.6$ \\
00032613079 & 0.32 & 2015-08-24.020 & 57258.0195 & -4.260 & $<1.9$ & $<4.6$ \\
00032613080 & 0.64 & 2015-08-24.340 & 57258.3398 & -3.940 & $<1.2$ & $<2.8$ \\
00032613081 & 0.60 & 2015-08-24.539 & 57258.5391 & -3.741 & $<1.3$ & $<3.2$ \\
00032613082 & 0.49 & 2015-08-24.805 & 57258.8047 & -3.475 & $<1.3$ & $<3.2$ \\
00032613083 & 0.57 & 2015-08-25.016 & 57259.0156 & -3.264 & $<1.3$ & $<3.2$ \\
00032613084 & 0.59 & 2015-08-25.340 & 57259.3398 & -2.940 & $<1.6$ & $<3.4$ \\
00032613085 & 0.46 & 2015-08-25.559 & 57259.5586 & -2.721 & $<1.4$ & $<3.3$ \\
00032613086 & 0.47 & 2015-08-25.887 & 57259.8867 & -2.393 & $<1.3$ & $<3.1$ \\
00032613087 & 0.58 & 2015-08-26.012 & 57260.0117 & -2.268 & $<1.3$ & $<3.2$ \\
00032613088 & 0.49 & 2015-08-26.406 & 57260.4062 & -1.874 & $<1.3$ & $<3.1$ \\
00032613089 & 0.53 & 2015-08-26.613 & 57260.6133 & -1.667 & $<1.2$ & $<2.9$ \\
00032613090 & 0.51 & 2015-08-26.801 & 57260.8008 & -1.479 & $<1.7$ & $<4.1$ \\
00032613091 & 0.85 & 2015-08-27.012 & 57261.0117 & -1.268 & $<1.0$ & $<2.5$ \\
00032613092 & 0.79 & 2015-08-27.406 & 57261.4062 & -0.874 & $<1.5$ & $<3.2$ \\
00032613093 & 0.67 & 2015-08-27.602 & 57261.6016 & -0.678 & $<1.3$ & $<3.1$ \\
00032613094 & 0.87 & 2015-08-27.801 & 57261.8008 & -0.479 & $<1.0$ & $<2.4$ \\
00032613095 & 0.56 & 2015-08-27.012 & 57261.0117 & -1.268 & $<1.2$ & $<3.0$ \\
00032613096 & 0.74 & 2015-08-28.008 & 57262.0078 & -0.272 & $<1.2$ & $<2.9$ \\
00032613097 & 0.80 & 2015-08-28.406 & 57262.4062 & 0.126 & $<1.1$ & $<2.7$ \\
00032613098 & 0.67 & 2015-08-28.602 & 57262.6016 & 0.322 & $<1.2$ & $<3.0$ \\
00032613099 & 0.87 & 2015-08-28.801 & 57262.8008 & 0.521 & $<1.2$ & $<2.5$ \\
00032613100 & 0.87 & 2015-08-29.004 & 57263.0039 & 0.724 & $<1.0$ & $<2.4$
\enddata
\tablenotetext{a}{Dead-time corrected exposure time.}
\tablenotetext{b}{Start date of the observation.}
\tablenotetext{c}{Time in days after the optical eruption of nova \nova 
on 2015-08-28.28 UT (MJD 57262.28).}
\end{deluxetable*}


\begin{deluxetable*}{rrrrrrr}
\tabletypesize{\scriptsize}
\tablecaption{\swift observations for the merged data
\label{tab:obs_merge}}
\tablewidth{0pt}
\tablehead{
\colhead{ObsID} & \colhead{Exp} & \colhead{Date\tablenotemark{a}} & \colhead{MJD\tablenotemark{a}} & \colhead{$\Delta t$\tablenotemark{a}} & \colhead{$3\sigma$ ulim} & \colhead{$5\sigma$ ulim}\\
\colhead{} & \colhead{[ks]} & \colhead{[UT]} & \colhead{[d]} & \colhead{[d]} & \colhead{[\power{-2} ct s$^{-1}$]} & \colhead{[\power{-2} ct s$^{-1}$]} }
\startdata
00032613063/064 & 1.13 & 2015-08-20.15 & 57254.1542 & -8.126 & $<0.7$ & $<1.6$ \\
00032613064/065 & 1.17 & 2015-08-20.41 & 57254.4140 & -7.866 & $<0.5$ & $<1.3$ \\
00032613065/066 & 1.20 & 2015-08-20.68 & 57254.6836 & -7.596 & $<0.5$ & $<1.2$ \\
00032613066/067 & 0.90 & 2015-08-20.92 & 57254.9218 & -7.358 & $<0.7$ & $<1.6$ \\
00032613067/068 & 0.88 & 2015-08-21.15 & 57255.1523 & -7.128 & $<0.7$ & $<1.7$ \\
00032613068/069 & 1.01 & 2015-08-21.42 & 57255.4180 & -6.862 & $<0.8$ & $<1.8$ \\
00032613069/070 & 0.94 & 2015-08-21.68 & 57255.6836 & -6.596 & $<0.9$ & $<1.9$ \\
00032613070/071 & 0.96 & 2015-08-21.92 & 57255.9180 & -6.362 & $<0.7$ & $<1.6$ \\
00032613071/072 & 1.27 & 2015-08-22.18 & 57256.1836 & -6.096 & $<0.5$ & $<1.2$ \\
00032613072/073 & 1.26 & 2015-08-22.44 & 57256.4434 & -5.837 & $<0.5$ & $<1.3$ \\
00032613073/074 & 0.96 & 2015-08-22.68 & 57256.6778 & -5.602 & $<0.7$ & $<1.7$ \\
00032613074/075 & 0.96 & 2015-08-22.92 & 57256.9160 & -5.364 & $<0.6$ & $<1.5$ \\
00032613075/076 & 1.32 & 2015-08-23.18 & 57257.1816 & -5.098 & $<0.5$ & $<1.3$ \\
00032613076/077 & 1.23 & 2015-08-23.44 & 57257.4434 & -4.837 & $<0.6$ & $<1.4$ \\
00032613077/078 & 0.96 & 2015-08-23.68 & 57257.6758 & -4.604 & $<0.7$ & $<1.7$ \\
00032613078/079 & 0.83 & 2015-08-23.91 & 57257.9140 & -4.366 & $<0.8$ & $<2.0$ \\
00032613079/080 & 0.96 & 2015-08-24.18 & 57258.1797 & -4.100 & $<0.7$ & $<1.7$ \\
00032613080/081 & 1.24 & 2015-08-24.44 & 57258.4395 & -3.841 & $<0.6$ & $<1.5$ \\
00032613081/082 & 1.09 & 2015-08-24.67 & 57258.6719 & -3.608 & $<0.7$ & $<1.6$ \\
00032613082/083 & 1.06 & 2015-08-24.91 & 57258.9101 & -3.370 & $<0.7$ & $<1.6$ \\
00032613083/084 & 1.16 & 2015-08-25.18 & 57259.1777 & -3.102 & $<0.8$ & $<1.8$ \\
00032613084/085 & 1.05 & 2015-08-25.45 & 57259.4492 & -2.831 & $<0.8$ & $<1.8$ \\
00032613085/086 & 0.93 & 2015-08-25.72 & 57259.7226 & -2.557 & $<0.7$ & $<1.6$ \\
00032613086/087 & 1.05 & 2015-08-25.95 & 57259.9492 & -2.331 & $<0.6$ & $<1.6$ \\
00032613087/088 & 1.07 & 2015-08-26.21 & 57260.2090 & -2.071 & $<0.6$ & $<1.6$ \\
00032613088/089 & 1.02 & 2015-08-26.51 & 57260.5097 & -1.770 & $<0.6$ & $<1.5$ \\
00032613089/090 & 1.04 & 2015-08-26.71 & 57260.7070 & -1.573 & $<0.7$ & $<1.7$ \\
00032613090/091 & 1.36 & 2015-08-26.91 & 57260.9062 & -1.374 & $<0.6$ & $<1.5$ \\
00032613091/092 & 1.64 & 2015-08-27.21 & 57261.2090 & -1.071 & $<0.7$ & $<1.5$ \\
00032613092/093 & 1.46 & 2015-08-27.50 & 57261.5039 & -0.776 & $<0.8$ & $<1.7$ \\
00032613093/094 & 1.54 & 2015-08-27.70 & 57261.7012 & -0.579 & $<0.6$ & $<1.4$ \\
00032613094/096 & 1.61 & 2015-08-27.90 & 57261.9043 & -0.376 & $<0.5$ & $<1.3$ \\
00032613096/097 & 1.54 & 2015-08-28.21 & 57262.2070 & -0.073 & $<0.6$ & $<1.4$ \\
00032613097/098 & 1.47 & 2015-08-28.50 & 57262.5039 & 0.224 & $<0.6$ & $<1.4$ \\
00032613098/099 & 1.54 & 2015-08-28.70 & 57262.7012 & 0.421 & $<0.7$ & $<1.5$ \\
00032613099/100 & 1.74 & 2015-08-28.90 & 57262.9024 & 0.622 & $<0.6$ & $<1.3$
\enddata
\tablenotetext{a}{Midpoint between the two observations.}
\end{deluxetable*}

The strict $5\sigma$ limits indicate that we should have seen the X-ray 
flash if it had occurred with the predicted luminosity and spectrum 
during the time of the monitoring. The restrictive $3\sigma$ 
limits can be used to constrain the X-ray flux in a meaningful way. 
The corresponding data are given in Tables\,\ref{tab:obs} 
and \ref{tab:obs_merge}.

\section{IMPLICATION OF NON-DETECTION OF X-RAY/UV FLUXES}
\label{section_implication}

We have confirmed theoretically that the X-ray flash should 
last 14 hours (0.59 days) or longer (in Section \ref{section_model}). 
The X-ray flash was not, however, detected in our six-hour-cadence
eight-day observations preceding the 2015 outburst. 
In this section we examine two possible reasons for the non-detection; 
(1) the X-ray flash had occurred during the \swift observation period,
but all the photons were obscured by surrounding neutral hydrogen, 
or (2) the X-ray flash had occurred earlier than our \swift 
observation period, i.e., more than eight days before the optical discovery.

\subsection{Absorption by surrounding neutral hydrogen}
\label{section_implication_abs}

A WD in a binary is possibly surrounded by ionized/neutral material
originating from the companion star. 
If the WD is surrounded by a substantial amount of neutral hydrogen, 
X-ray photons emitted from the WD surface could be mostly absorbed,
and thus one may not detect the X-ray flash.  
It is, however, poorly known whether the mass-accreting WDs
in recurrent novae are surrounded 
by ionized or neutral matter in their early outburst phase.

The companion of M31N 2008-12a has not been identified, yet.
If the companion star is a Roche-lobe filling subgiant, we can expect 
that the mass transfer is mainly through the accretion disk and 
a small proportion of the mass lost by the donor 
is spread over the circumbinary region. If the 
companion is a red giant, the binary could be embedded within the cool neutral wind, 
which absorbs supersoft X-ray photons from the WD. 

\citet{dar14} compared the spectral energy distribution (SED) 
of M31N 2008-12a in its quiescent phase with 
those of the Galactic recurrent novae, RS~Oph, T~CrB, and U~Sco. 
Based on the similarity of the RS~Oph SED, rather than U~Sco which is 
much fainter, the authors suggested that M31N 2008-12a likely contains 
a red giant companion with a significant accretion disk 
component that dominates the near-UV and optical flux. 
The authors note, however, that the possibility of 
a face-on subgiant companion remains 
because U~Sco is an eclipsing binary and its edge-on disk may not be bright.

\citet{hac16II} classified 40 classical novae into six classes according to 
their evolutionary path in the color-magnitude diagram and found that 
the different paths correspond to differences in the nova speed class 
and thus the envelope mass.  These authors also displayed 
the color-magnitude evolution during the 2014 outburst
of M31N 2008-12a and found that its characteristic properties are similar to 
those of U Sco and CI Aql, 
which are both recurrent novae with a subgiant companion,
but different from RS Oph which has a red giant companion
(see their Figures 72(c), (d), and 76(b)).
This suggests that M31N 2008-12a has a subgiant companion. 

The Galactic object RS~Oph is a well observed recurrent nova. 
Its recorded outbursts were in 1898, 1933, 1958, 1967, 1985 and 2006 \citep{eva08}. 
In the 1985 outburst very soft X-ray emission was detected 
251 days after the optical maximum \citep{mas87}. 
\citet{hac01} regarded this X-ray emission to be due to the accretion luminosity 
and suggested that the accretion rate had dropped by a factor of six after the 
outburst. \citet{dob94} also pointed out a decrease 
in the mass accretion rate from line fluxes in \ion{H}{1} and \ion{He}{1} 
that decreased by a factor of four after the 1985 outburst.
Day 251 falls in the period of the postoutburst minimum of 
days $\sim 100$-400, after which the visual luminosity increased 
by about a magnitude \citep{eva88}.
X-rays were also observed in 1992 with ROSAT \citep{ori93},
but the supersoft flux ($< 0.5$ keV) was very weak. 
One possible explanation is absorption by the massive cool wind
\citep[e.g.,][]{sho96} from the red giant companion as suggested by \citet{anu99}. 
After 21 years of accumulation, the overlying RG wind reaches 
2-5 $\times 10^{22} $cm$^{-2}$ \citep{bode06,sok06}  
in the 2006 outburst. However, the absorption effect of this overlying RG wind
is quickly removed \citep{osb11}.  After the outburst 
the mass-accretion rate had dropped 
in the post-outburst minimum phase and soft X-rays were observable
because the ejecta swept away the red giant cool wind. 
After day 400, the mass transfer had recovered and the hot component 
could be surrounded by neutral hydrogen.  

If the accretion disk is completely blown off by the ejecta, it may take 
a few orbital periods until a significant amount of the red giant wind 
falls onto the WD. 
\citet{hac01} roughly estimated the resumption time of mass-transfer 
in RS Oph to be
$\Delta t = a / v \sim 300 R_\sun / 10 \mbox{~km~s}^{-1}= 300$ days, 
where $a$ is the binary separation and $v$ is the velocity of infalling matter.  
This is roughly consistent with the recovery of the quiescent $V$ 
luminosity 400 days after the 1985 outburst [see Figure 1 in \citet{eva88};
Figure 2 in \citet{hac06kb}, \citet{wor07}, also \citet{dar08}].
M31N 2008-12a shows a ultra-short recurrence period of one or half a year.  
It is unlikely that systems like RS~Oph produce successive outbursts 
with such a short recurrence period because of a long interruption 
of mass transfer unless the disk survives the eruption. 
Thus, we expect that M31N 2008-12a does not have a red giant companion.  

U~Sco is another well observed Galactic recurrent nova with a subgiant companion. 
\citet{ness12} examined an X-ray eclipse during the 2010 outburst in detail 
and concluded that the mass accretion resumed as early as day 22.9, 
midway during the SSS phase. 
In a binary with a subgiant companion, thus, we can expect 
the mass accretion to resume just after an outburst. 

For these reasons, 
we may conclude that M31N 2008-12a has a subgiant companion.
In case of close binaries, the transferred matter is mostly distributed  
in the orbital plane \citep[see, e.g.,][for a 3-D calculation of mass flow
in a close binary]{syt09}. 
Note that, in our binary models, 
the WD radiates $L_{\rm ph}=200$--$500~L_\sun$ 
at $T_{\rm ph}=4$--$5 \times 10^5$ K in its quiescent phase. 
Therefore, we expect that the matter surrounding the WD may be kept 
ionized during the quiescent phase.
Thus, we consider that an X-ray flash should have been detected
if it had occurred during our observing period.

\subsection{Slow evolution after X-ray flash}
\label{section_implication_slowevolution}

The other explanation of the undetected X-ray flash is that 
the flash had already occurred and finished when we started  
our observations eight days before the UV/optical discovery. 
This means that the evolution time from C to D in Figure \ref{hr}
was longer than eight days, and 
the optical/UV bright phase, from D to E, lasted 
about 5.5 days \citep{dar15,hen15}. 
\citet{dar15} pointed out that 
M31N 2008-12a showed slow rise to the optical peak 
magnitude in the 2014 outburst. This suggest a 
slow evolution toward point D.

The timescale from C to E can be roughly estimated as follows. 
The decrease of the envelope mass from C to E is owed both 
to nuclear burning and mass ejection. 
For example, in a $1.38~M_\sun$ WD with $P_{\rm rec}=0.47$ yr,
the envelope mass is $1.5\times 10^{-7}~M_\sun$ at 
C and decreases to $7.5\times 10^{-8}~M_\sun$ at E.
In the 2014 outburst, the ejected hydrogen mass was estimated to be  
$M_{\rm ej,H}=(2.6 \pm 0.4)\times 10^{-8}~M_\sun$ \citep{hen15},
which corresponds to $M_{\rm ej} =4.7 \times 10^{-8}~M_\sun$ for $X=0.55$
(the hydrogen mass fraction is smaller than the initial 
$X=0.7$ because of convective mixing with the nuclear burning region).
 
\citet{dar15} derived a total ejected mass of 
$\gtrsim 3 \times 10^{-8} M_\sun$. 
Here we assume the mass ejected by the wind to be 
$M_{\rm ej} =4.7 \times 10^{-8}~M_\sun$.
Thus, nuclear burning had consumed the rest of the mass,
$\Delta M_{\rm env}=1.5\times 10^{-7}~M_\sun - 7.5
\times 10^{-8}~M_\sun-4.7 \times 10^{-8}~M_\sun
= 3.1 \times 10^{-8}~M_\sun$ during the period from C to E. 
If we take the mean nuclear luminosity as $\log L_{\rm nuc}/L_\sun=$4.65, 
the evolution time from C to E is roughly estimated as 
$(\Delta M_{\rm env}\times X\times \epsilon_{\rm H})/
L_{\rm nuc}=14.8$ days, here $X=0.55$ and energy generation
of hydrogen burning $\epsilon_{\rm H}=6.4\times 10^{18}$erg~g$^{-1}$. 
So we obtain the duration between epochs C and D in Figure \ref{hr} to be
$14.8-5.5=9.3$ days.   For a longer recurrence period, 
$P_{\rm rec}=0.95$ yr, we obtain, in the same way,  
($1.9 \times 10^{-7}~M_\sun -7.3\times 10^{-8}~M_\sun-4.7
\times 10^{-8}~M_\sun) X \epsilon_{\rm H}/L_{\rm nuc}-5.5=21 -5.5 =15.5$ days.
Here we assume X=0.53 and the mean nuclear luminosity to be 
$\log L_{\rm nuc}/L_\sun=$4.85 for a somewhat stronger 
shell flash than in the shorter recurrence period 
(see Figure \ref{L.138}).

In this way we may explain that the X-ray flash had occurred 15.5 days 
($P_{\rm rec}=0.95$ yr) or 9.3 days ($P_{\rm rec}=0.47$ yr) before the 
optical/UV peak.  These values should be considered as rough estimates 
because they are sensitive to our simplified value for the mean $L_{\rm nuc}$, 
beside other parameters such as the WD mass and recurrence period 
(i.e., mass accretion rate). 
Even though, these estimates suggest that the nova evolution is slow between 
C and D, and the X-ray flash could have occurred 
before our observing period (eight days before the optical/UV detection),
rather than immediately before the optical maximum \citep{kat15sh}. 

Observations of recurrent novae have shown that they evolve much faster 
than typical classical novae, which is demonstrated by their 
very short X-ray turn-on time \citep[duration between the optical peak to 
the X-ray turn-on, see, e.g.,][for the shortest 4 day case of 
V745~Sco]{pag15}. 
By analogy, and without observational support, we 
suspect that the rising phase of recurrent novae must also be fast. 
However, our 8-days-non-detection prior to 
the optical/UV peak suggests it may not be as fast as predicted. 
It is partly suggested by our calculation that $L_{\rm nuc}^{\rm max}$ is 
rather small in very short recurrence period novae even though the WD 
is extremely massive. The small nuclear burning energy generation
rate renders the recurrent nova eruption relatively weak. 
Because time-dependent calculations have many 
difficulties in the expanding phase of nova outbursts, no one has ever
succeeded in reproducing reliable multiwavelength light curves that
included the rising phase. 
We expect that the detection of X-ray flashes can confirm such 
a slow evolution in the very early phase of a nova outburst.

\section{DISCUSSION}\label{section_discussion} 
 
\subsection{Comparison with other works}
Many numerical calculations of shell flashes have been presented, but only 
a few of them provided sufficient information on 
the early stages corresponding to the X-ray flash. 
\citet{nar80} calculated hydrogen shell flashes, in which  
the evolution time from $L_{\rm nuc}^{\rm max}$ (defined as $t=0$) to 
a stage of $\log T_{\rm ph}$ (K)$\sim 5.45$ is 1.5 hours
for a 1.3 $M_\sun$ WD with 
$\dot M_{\rm acc}=1 \times 10^{-10} M_\sun$ yr$^{-1}$. 
\citet{ibe82} showed the timescale from $t=0$ to $\log T_{\rm ph}$ (K) 
$=5.5$ to be about 100 days for a 0.964 $M_\sun$ WD with 
$\dot M_{\rm acc}=1.5 \times 10^{-8}M_\sun$ yr$^{-1}$ (in his Figure 8). 
For a 1.01 $M_\sun$ WD, it is  
about 20 days with $\dot M_{\rm acc}=1.5 \times 10^{-8}M_\sun$ yr$^{-1}$   
and about 1 day for $\dot M_{\rm acc}=1.5 \times 10^{-9}M_\sun$ yr$^{-1}$ 
(in his Figures 15 and 21) . 
These studies are based on the Los Alamos opacities that have no 
Fe peak, so the optically thick wind does not occur, 
resulting in a much longer total duration of the nova outburst. 
However, the timescales in the very early phase 
corresponding to the X-ray flash (defined by $\log T_{\rm ph}$ (K) $>5.6$) 
should not be much affected by the Fe peak
because the photospheric temperature is higher than the Fe peak. 
We see a tendency of a longer X-ray flash for a less massive WD  
and for a larger mass accretion rate (i.e., a shorter recurrence period).  
This tendency agrees with our results. 

\citet{hil14} showed evolutionary change in the effective temperature of 
nova outbursts with the OPAL opacities. Their Figure 3 shows 
an X-ray flash duration (defined by $\log T_{\rm eff}$ (K) $> 5.5$)
of a few hours for 
a 1.4 $M_\sun$ WD with a mass accretion rate of $10^{-8}M_\sun$ yr$^{-1}$. 
This accretion rate corresponds to the recurrence period of 20 yr 
\citep{pri95}.  Considering the tendency that a longer duration 
X-ray flash is obtained for a less massive WD and larger 
mass-accretion rate, their duration of a few hours is consistent 
with our results of half a day to one day (Table \ref{table_models}). 

\subsection{General relativistic stability of massive WDs}
\label{sec_stability}
The masses of our WD models are very close to the Chandrasekhar mass limit, 
above which non-rotating WDs cannot exist.
This limit is $1.457~(\mu_{\rm e}/2)^{-2}~M_\sun$ for pure degenerate gas 
\citep[e.g. Equation 6.10.26 in][]{sha83}, 
where $\mu_{\rm e}$ is the mean molecular weight of the electron. 
According to \citet{sha83}, \citet{kap49} first pointed out
that general relativity
probably induces a dynamical instability when the radius of a WD becomes
smaller than $1.1 \times 10^3$ km. 
\citet{cha64} independently showed that a WD of mass 
$> 1.4176~M_\sun$ is dynamically unstable when its radius decreases 
below $1.0267 \times 10^3$ km. 
This means that the instability occurs at radii much larger 
than the Schwarzschild radius $R_{\rm S}=2GM/{c^2}$. 
In our model, the 1.38~$M_\sun$ WD has the radius of $\sim 2000$ km, 
much larger than the Schwarzschild radius 
$2G(1.38M_\sun)/{c^2}=$4.1 km, and the above stability limits
of general relativity, $R_{\rm GR} \sim 1000$ km.


\begin{figure}
\epsscale{1.20}
\plotone{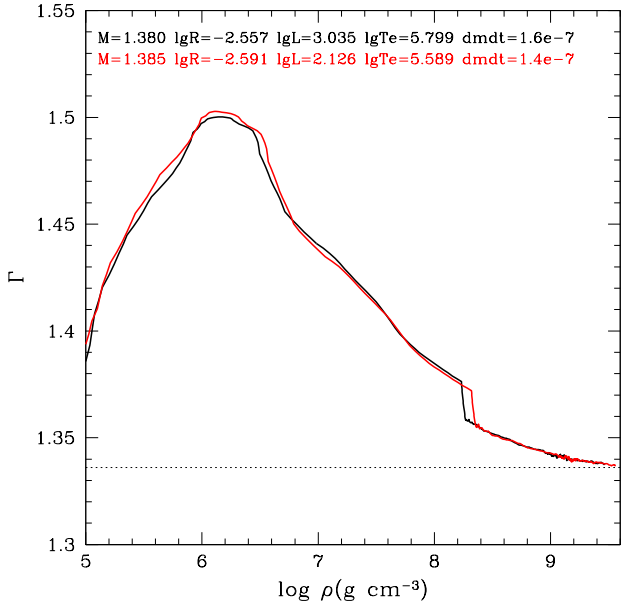}
\caption{Distribution of $\Gamma=d \log P / d \log \rho$
in our models of 1.38 $M_\sun$ (black) and $1.385~M_\sun$ (red). 
The rightmost point corresponds to the center of the WD. 
The horizontal dotted line denotes the stability line of $\Gamma=1.336$.
See Section \ref{sec_stability} for more detail.   
}
\label{fig:gamma}
\end{figure}

Assuming the polytropic relation $P\propto \rho^\Gamma$, 
\citet{sha83} derived a different stability criterion 
(see their Equation 6.10.30), i.e.,
\begin{equation}
\Gamma - {4\over 3} = 1.125 ({2GM\over{R c^2}}). 
\label{general_relativity_stability}
\end{equation}
In our 1.38~$M_\sun$ model with $\dot M_{\rm acc}=1.6 
\times 10^{-7}~M_\sun$~yr$^{-1}$ ($P_{\rm rec}=0.95$ yr), 
the right hand side of Equation (\ref{general_relativity_stability})
becomes maximum at point A in Figure \ref{hr}, that is,
$1.125 \times 0.00215 = 0.00242$. 
Thus, the stability criterion becomes $ \Gamma > 1.3358$. 
For the 1.385 $M_\sun$ model with $1.4 \times 10^{-7}~M_\sun$~yr$^{-1}$ 
($P_{\rm rec}=0.97$ yr), this criterion is $\Gamma > 1.3359$, 
essentially the same as for the 1.38 $M_\sun$ model.
We calculated the distribution of $\Gamma$ in the accreting phase 
as shown in Figure \ref{fig:gamma}.
The black line depicts $\Gamma= d \log P / d \log \rho$ 
of the 1.38~$M_\sun$ model, while 
the red line represents the 1.385~$M_\sun$ model. 
Both the red and black lines are located above 
the horizontal dashed line of $\Gamma=1.336$. 
Therefore, both models satisfy the stability condition ($\Gamma>1.336$).
Note that the central part hardly changes during the flash. 
Thus, we conclude that our 1.38 and 1.385 $M_\sun$ WD models are stable
against the general relativistic instability.

\subsection{The soft X-ray transient MAXI J0158$-$744}

MAXI J0158$-$744 is an X-ray transient, believed to be a Be star plus 
WD binary that appeared in the Small Magellanic Cloud \citep{li12,mor13}. 
MAXI detected a brief ($ <$ 90 min) X-ray flux ($ <$ 5 keV)   
of very high luminosity (several $\times 10^{39}-10^{40}$ erg~s$^{-1}$).
Follow up \swift observations detected soft X-ray emission 
($\sim 100$ eV) that lasted two weeks \citep{li12}, 
resembling a SSS phase on a massive WD. 
For the origin of the early brief X-ray flux, 
\citet{li12} attributed to the interaction of 
the ejected nova shell with the Be star wind. 
\citet{mor13} concluded that the X-ray emission is unlikely to have 
a shock origin, but associated it with the fireball stage of a nova outburst 
on an extremely massive WD. 

In this paper we have considered the very early phase of shell flashes 
in the extreme limit of massive WDs and high mass-accretion rates. 
Our calculations have shown that, in this limit, 
the evolution is very slow (X-ray flash lasts $\sim$one day), 
and the X-ray luminosity does not exceed the Eddington luminosity 
($\sim 2 \times 10^{38}$ erg~s$^{-1}$). 
The wind mass loss does not occur during the X-ray flash, 
and the energy range of the X-ray photons are up to 100-120 eV.  
These properties are incompatible with the bright (super-Eddington),
high energy ($< 5$ keV), very short duration ($<$90 min)  
X-ray emission seen early on in MAXI J0158-744. 
Therefore, we conclude that the brief early X-ray flux 
in MAXI J0158-744 is not associated with that expected in 
the extreme limit of massive
non-rotating WDs with high mass-accretion rates.

\section{CONCLUSIONS}\label{section_conclusion}

Our main results are summarized as follows.

\noindent
1. In a very early phase of a recurrent nova outburst, the photospheric 
luminosity rises very close to the Eddington luminosity at the photosphere 
and the temperature reaches as high as $T_{\rm ph}\sim 10^6$ K 
in WDs as massive as $1.38~M_\odot$.   
We expect bright supersoft X-ray luminosities in this X-ray flash phase, 
as large as $L_{\rm X} \sim 10^{38}$ erg~s$^{-1}$. 

\noindent
2. 
We present light curves of X-ray flashes for 1.35, 1.38, 
and $1.385 ~M_\odot$ WDs.  The duration of the X-ray flash depends 
on the WD mass and the recurrence period, shorter 
for a more massive WD, and longer for a shorter recurrence period. 
The duration of the X-ray flash would be a good indicator of the WD mass and 
mass-accretion rate because it depends sensitively on these values.

\noindent
3. The optically thick wind arises at the end of X-ray flash 
($\log T_{\rm ph}$ (K) $\sim 5.6$) owing to acceleration 
by the Fe opacity peak.
As no strong wind mass loss is expected during the X-ray flash, 
we could observe a naked photosphere, i.e., the spectrum is close to
that of blackbody with $T_{\rm ph}$.

\noindent
4. We observed with a six-hour-cadence the 2015 outburst of 
M31N 2008-12a with \swift from eight days before the optical discovery. 
Although our theoretical prediction of the X-ray flash duration 
was long enough, as long as 0.5 -- 1.5 days,  no X-ray flash was detected.

\noindent
5. We examined two possible reasons for the non-detection. 
Absorption by the surrounding 
matter originated from the companion is unlikely. Instead, 
we suggest that the X-ray flash could have occurred 
before our observations started, because short recurrence period novae
undergo a very slow evolution.

\noindent
6. The X-ray flash is one of the last frontiers of nova studies. 
We encourage further attempts at observational confirmation in the near future.  Any detection of
X-ray flashes would be essentially important to explore the pre-optical-maximum
phase and to ultimately understand the complete picture of nova eruptions.


\acknowledgments
We are grateful to the \swift Team for the excellent scheduling of the 
observations, in particular the duty scientists and the science planners.
We also thank the anonymous referee for useful comments 
that improved the manuscript.
This research has been supported in part by Grants-in-Aid for
Scientific Research (24540227, 15K05026, 16K05289) 
of the Japan Society for the Promotion of Science. M. Henze acknowledges 
the support of the Spanish Ministry of Economy and Competitiveness (MINECO) 
under the grant FDPI-2013-16933. 
JPO and KLP acknowledge support from the UK Space Agency.
AWS thanks the NSF for support through AST1009566.
M. Hernanz acknowledges MINECO support 
under the grant ESP2014-56003-R.

%



%
%


%
%











\end{document}